\documentclass[aip,jcp,10pt,twocolumn,amsmath,amssymb]{revtex4-1}
\usepackage{graphicx}
\usepackage{epstopdf}
\usepackage[utf8x]{inputenc}
\usepackage{amsmath}
\usepackage{url}
\usepackage{hyperref}

\usepackage{epsfig}
\usepackage{color}

\begin{document}

\title{Computing the phase diagram of binary mixtures: a patchy particle case study}

\author{Lorenzo Rovigatti} 
\affiliation{{Dipartimento di Fisica, Universit\`a di Roma {\em La Sapienza}, Piazzale A. Moro 2, 00185 Roma, Italy} }
\author{Daniel de las Heras}
\affiliation{Theoretische Physik II, Physikalisches Institut,
  Universit{\"a}t Bayreuth, D-95440 Bayreuth, Germany}
\author{Jos\'e Maria Tavares}
\affiliation{Instituto Superior de Engenharia de Lisboa ISEL,
Rua Conselheiro Em\'{\i}dio Navarro 1, P-1950-062 Lisbon, Portugal }
\affiliation{
Centro de F\'{\i}sica Te\'{o}rica e Computacional da Universidade de Lisboa,
Avenida Professor Gama Pinto 2, P-1649-003 Lisbon, Portugal
}
\author{Margarida M. Telo da Gama}
\affiliation{Departamento de F\'{i}sica, Faculdade de Ci\^{e}ncias da Universidade de Lisboa, Campo Grande, P-1749-016 Lisbon, Portugal and Centro de F\'{i}sica Te\'{o}rica e Computacional da Universidade de Lisboa, Avenida Professor Gama Pinto 2, P-1649-003 Lisbon, Portugal}
\author{Francesco Sciortino} 
\affiliation{{Dipartimento di Fisica and  CNR-ISC, Universit\`a di Roma {\em La Sapienza}, Piazzale A. Moro 2, 00185 Roma, Italy} }

\begin{abstract}
We investigate the phase behaviour of 2D mixtures of bi-functional and three-functional patchy particles and 3D mixtures of bi-functional and tetra-functional patchy particles by means of Monte Carlo simulations and Wertheim theory. We start by computing the critical points of the pure systems and then we investigate how the critical parameters change upon lowering the temperature. We extend the Successive Umbrella Sampling method to mixtures to make it possible to extract information about the phase behaviour of the system at a fixed temperature for the whole range of densities and compositions of interest.

\end{abstract}

\maketitle

\section{Introduction}

The role of the particle valence --- defined as the ability  to form only a limited number of bonds with neighbouring particles ---  in controlling the phase behaviour of colloidal systems has been emphasized in numerous recent studies~\cite{Zacca1, bian,lungo,FoffiKern,John}.
Limited valence has emerged as the key element in the formation of equilibrium colloidal networks, commonly named gels.  In this state of matter, rigidity is enforced by the long-life of
the inter-particle bonds, which, at low temperatures $T$, is longer than the experimental
observation time~\cite{John,rovigatti_molphys,lr_dos}. Lowering the valence guarantees that these low $T$ gel states
do not clash with phase separation, which is always present when the thermal energy becomes
significantly lower than the characteristic bond strength.
In contrast with the out-of-equilibrium colloidal gels formed via spinodal decomposition
followed by a kinetic arrest induced by the strong depletion attraction~\cite{LuCiulla}, limited valence
gels are equilibrium states.  It has indeed been shown that
lowering the valence shifts the phase separation boundaries to low densities~\cite{Zacca1,bian}, opening a wide region of particle concentrations where stable gels can form~\cite{laponite_nat_mat}.

Binary mixtures of limited valence particles enrich considerably more the spectrum of possibilities offered by limited valence.
First, the average valence  of the system can take non-integer values, which has been exploited to investigate the approach to the limit of valence two. In this limiting case,
phase separation is completely suppressed, since  particles aggregate in long chains that interact only through excluded volume interactions. Second, mixtures can be exploited to
tune the selectivity of the network to different species, stabilizing
mixed or interpenetrating gels~\cite{goyal_bicontinuous_gels,delasheras_bicontinuous_gels}, the equilibrium equivalent of the recently reported
out-of-equilibrium bigels~\cite{varrato2012arrested}.

In limited valence systems, gas-liquid phase separation arises from a subtle competition
between the number of bonds that can form in the two coexisting phases (the energy term) and the  entropy. The latter accounts for the degeneracy of the bonding patterns, which is  different in the two phases. The gas phase is usually formed by diluted clusters, while the liquid phase
is characterized by a percolating network of bonds.  This competition is captured by the thermodynamic perturbation theory (TPT) developed by Wertheim~\cite{wertheim1,wertheim2,wertheim3,wertheim4} to model the behaviour of associating fluids, the atomic and molecular analogues of
limited-valence colloids.  Wertheim theory is a
powerful tool for investigating the phase behaviour of pure fluids as well as of binary mixtures~\cite{mixtures_lisbona_1,mixtures_lisbona_2}.

Recent theoretical studies of binary mixtures with different compositions have revealed a subtle interplay between the entropy of bonding and the entropy of mixing, with a marked effect on the phase diagram of the mixture. Interestingly, Wertheim theory predicts that for binary mixtures of bi- and three-functional particles, the gas-liquid critical density  $\rho_c$ and the critical temperature $T_c$ decrease as the average valence decreases ({\it i.e.,} on increasing the fraction of bi-functional
particles)~\cite{mixtures_lisbona_1}, in agreement with existing numerical results~\cite{bian}. By contrast, for a binary mixtures of bi- and tetra-functional particles,  Wertheim theory
predicts a qualitative difference in the behaviour of the critical parameters. Indeed, binary mixtures of such species are expected to initially undergo an increase of $\rho_c$ as the average valence decreases. A further increase in the number of bi-functional particles inverts this trend and $\rho_c$ decreases and tends to a constant value as the fraction of bi-functional particles approaches one. 
Finally, mixtures of two and five functional particles are predicted to have critical densities which increase monotonically as the average valence is lowered~\cite{mixtures_lisbona_1}.

The numerical evaluation of
 phase coexistence in binary mixtures is not an easy task. Indeed, the presence of a
second species  adds a new axis to the phase diagram, which is now a three-dimensional volume defined by $T$ and the density of the two components ($T-\rho_A-\rho_B$)  or by other  combinations of $\rho_A$ and $\rho_B$
as $T-\rho-x$, where $\rho=\rho_A+\rho_B$ is the total number density and $x=\rho_B/\rho$ is the composition of the mixture~\cite{rowlinson1959liquids,wilding_mixture}.  The two coexisting phases
are characterized by different values of $x$.  Even focusing on a specific $x$ value (along
the so-called dilution line), the determination of the phase diagram requires the evaluation
of the shadow lines~\cite{sollich_poly_review}.
In this manuscript we  introduce a new and powerful computational method to investigate the phase behaviour of mixtures in the whole three-dimensional volume, by extending the successive umbrella sampling method~\cite{sus}, which was shown to be
very effective in the evaluation of the phase behaviour of single component systems~\cite{virnau_alkanes_mixtures,schilling_sus,prl-lisbona,dhs_prl,gallo_prl_sus}. By applying this new methodology it is possible to
compute the entire density of states for the binary mixture, {\it i.e.,} the information
required to estimate phase coexistence at all compositions.
As a test case of scientific relevance we compute the phase diagram of mixtures of bi-functional and three-functional patchy particles  in two-dimensions ($2D$) and of bi-functional and tetra-functional patchy particles in three dimensions ($3D$).  We then compare numerical results with theoretical predictions based on Wertheim theory, confirming the predicted growth of the critical density in the $2-4$ mixture as the fraction of bi-functional particles is increased. The analysis of the calculated phase behaviour shows that the $\rho_c$ growth on increasing the number of bi-functional particles results from a progressive transformation of the transition from condensation to demixing.  At the same time, increasing the number of particles with two patches does reduce the region in density where the instability takes place, confirming the general trend that a reduction of the average valence increases the density region where a stable gel can form.

\section{Methods}

\subsection{Model}

Each particle is modelled as a hard sphere (in 3D) or a hard disk (in 2D) of diameter $\sigma$, decorated with a fixed number of interacting patches on the surface. The patch-patch interaction between particles $i$ and $j$ is described by a Kern-Frenkel (KF) potential, {\it i.e.,} it is a square-well potential of range $\delta$ and depth $\epsilon$, modulated by a function $f(\hat{\boldsymbol\Omega}_i, \hat{\boldsymbol\Omega}_j)$ which depends solely on the particle orientations $\hat{\boldsymbol\Omega}_i$ and $\hat{\boldsymbol\Omega}_j$. Let $\mathbf{\hat{r}}_{ij}$ be the normalized vector joining the centres of particles $i$ and $j$ and $\mathbf{\hat{v}}_i^\alpha$ the versor connecting the centre of particle $i$ with the patch $\alpha$ on its surface. The function $f$ can then be written as

\begin{equation}
f(\hat{\boldsymbol\Omega}_i, \hat{\boldsymbol\Omega}_j) = \left\{ \begin{array}{rl}  
1 & \mathrm{if} \left\{ \begin{array}{rl} 
\mathbf{\hat{r}}_{ij} \cdot \mathbf{\hat{v}}_i^\alpha > \cos{\theta_\mathrm{max}} & \mathrm{for\; any\; \alpha},\\
\mathbf{\hat{r}}_{ji} \cdot \mathbf{\hat{v}}_j^\beta > \cos{\theta_\mathrm{max}} & \mathrm{for\; any\; \beta},
\end{array}\right.\\
0 & \mathrm{otherwise},
\end{array} \right.
\end{equation}

\noindent
where $\theta_\mathrm{max}$ controls the width of the patches. All the patches share the same shape, {\it i.e.,} $\delta$ and $\theta_\mathrm{max}$ are fixed and do not depend on particle species.

The only difference between particles of different species is in the number of patches patterning their surfaces. In 2D we study binary mixtures of particles decorated by either $3$ (species $A$) or $2$ (species $B$) patches~\cite{john_2D}. For particles of species $A$, the patches are symmetrically placed on the equator, for particles of species $B$ they are located on the poles. The KF parameters are $\delta = 0.03\sigma$ and $\cos{\theta_\mathrm{max}} = 0.894$. 

In 3D we study a binary mixture with the two species having either $4$ (species $A$) or $2$ (species $B$) patches. Patches are located on the particle surface in a tetrahedral fashion~\cite{Rom09a} for species-$A$ particles, on the poles for species-$B$ particles. The KF parameters are $\delta = 0.119\sigma$ and $\cos{\theta_\mathrm{max}} = 0.92$.

Both sets of parameters fulfill the geometrical single-bond-per-patch condition $\sin{\theta_\mathrm{max}} \leq 1 / (2(1+\delta))$,  preventing patches from being involved in more than one bond~\cite{FoffiKern}.

\subsection{Computational methods}

\subsubsection{Pure systems}

To compute the location of the (pseudo-)critical points of the pure systems we rely on the Bruce--Wilding (BW) mixed-field scaling method~\cite{bruce_wilding_energy_mixing}. This technique provides an expression for the order parameter $M$ which can be used to fit the probability distribution $P(M)$ of non-symmetric fluids to the Ising one, in order to estimate the pseudo-critical parameters of the finite system. On top of that, the BW approach provides scaling expressions which can be used to obtain the values of the critical parameters in the thermodynamic limit. With this procedure, the deviation from the average value of the order parameter $M$ at criticality can be written as~\cite{bruce_wilding_energy_mixing}

\begin{equation}
\label{eq:M_BW}
\Delta M = M - M_c \propto \rho + s u,
\end{equation}

\noindent
where $\rho = N/V$ is the number density, $N$ is the number of particles, $V$ is the volume of the system, $u = U/V$ is the energy density and $s$ is a non-universal ({\it i.e.,} model-dependent) factor.

In order to calculate the joint probability distribution $p(N, U)$, to be compared with the probability distribution of the Ising order parameter, we rely on Grand Canonical Monte Carlo (GMMC) simulations, {\it i.e.,} simulations at fixed temperature $T$, number of particles $N$ and chemical potential $\mu$. We employ successive umbrella sampling (SUS)~\cite{sus} to overcome the high free-energy barriers between the two phases.
With this method, the region to be explored is partitioned in  overlapping windows of  $\Delta N$ particles.
Each window is then sampled with GCMC simulations with appropriate boundary conditions~\cite{binder}, providing a
speed-up proportional to the number of windows explored in parallel. 

To properly locate the pseudo-critical point we make use of Eq.~\eqref{eq:M_BW} to project $p(N, U)$ and to obtain $p(\Delta M)$. In turn, we extract the critical parameters by matching this distribution to the (2D or 3D) Ising order parameter distribution~\cite{ising_2D,ising_3d} by means of histogram reweighting techniques~\cite{histrew}.

\subsubsection{Mixtures}
\label{subsubsec:mixtures}

Now we introduce this extended SUS method for the case of a generic mixture. Let $S$ be the number of species and $[N_i^{\mathrm{min}}, N_i^{\mathrm{max}}]$ be the range of number of particles of species $i$ of interest. Applying the SUS method consists in partitioning the $[N_1^{\mathrm{min}}, N_1^{\mathrm{max}}] \times [N_2^{\mathrm{min}}, N_2^{\mathrm{max}}] \times \ldots \times [N_S^{\mathrm{min}}, N_S^{\mathrm{max}}]$ space into overlapping windows of size $n_1 \times n_2 \times \ldots \times n_S$. Without any loss of generality we can fix the width of the overlap to be $\delta w$ and then the total number of windows is $n_w = \prod_{i=1}^S \lceil (N_i^{\mathrm{max}} - N_i^{\mathrm{min}}) / (n_i-\delta w)\rceil$, where $\lceil \cdot \rceil$ stands for the ceiling function.
Each window is identified by a $S$-dimensional index $w$. All the windows are then explored through special GCMC simulations, {\it i.e.,} simulations performed at fixed $T$, $V$, $\{\mu_i\}$, with the additional constraints that the number of particles of species $i$, $N_i$, has to lie within the range $[N_i^w, N_i^w + n_i]$ for each $i \in [1,S]$. The main simulation output is the histogram counting how many times a state with $N_1, N_2, \ldots, N_S$ has been visited in a given window $w$, namely $p^w(N_1, N_2, \ldots, N_S)$. The total free-energy density profile $p(N_1, N_2, \ldots, N_S)$ is then computed by using the overlapping portions of the windows to join together all the $p^w$. This operation is done through a least-squares method. Let $p^p(N_1, N_2, \ldots, N_s)$ be the partial, already joined part of the total $p(N_1, N_2, \ldots, N_S)$ and $p^w(N_1, N_2, \ldots, N_S)$ be the histogram of the next window to be attached. Then, in order to extend $p^p$ to the $\{N_i\}$ values stored in $p^w$ one needs to multiply the latter by the factor $b^w$ given by

\begin{equation}
\label{eq:join}
b^w = \frac{\sum_{{n_i} \in \{O^w\}} p^p(n_1, n_2, \ldots, n_S) p^w(n_1, n_2, \ldots, n_S)}{\sum_{{n_i} \in \{O^w\}} p^p(n_1, n_2, \ldots, n_S)},
\end{equation}
\noindent
where $\{O^w\}$ is the set of $N_1, N_2, \ldots, N_S$ values which are in the overlapping region between $p^p$ and $p^w$.

\begin{figure}
\centering \includegraphics[width=1\linewidth]{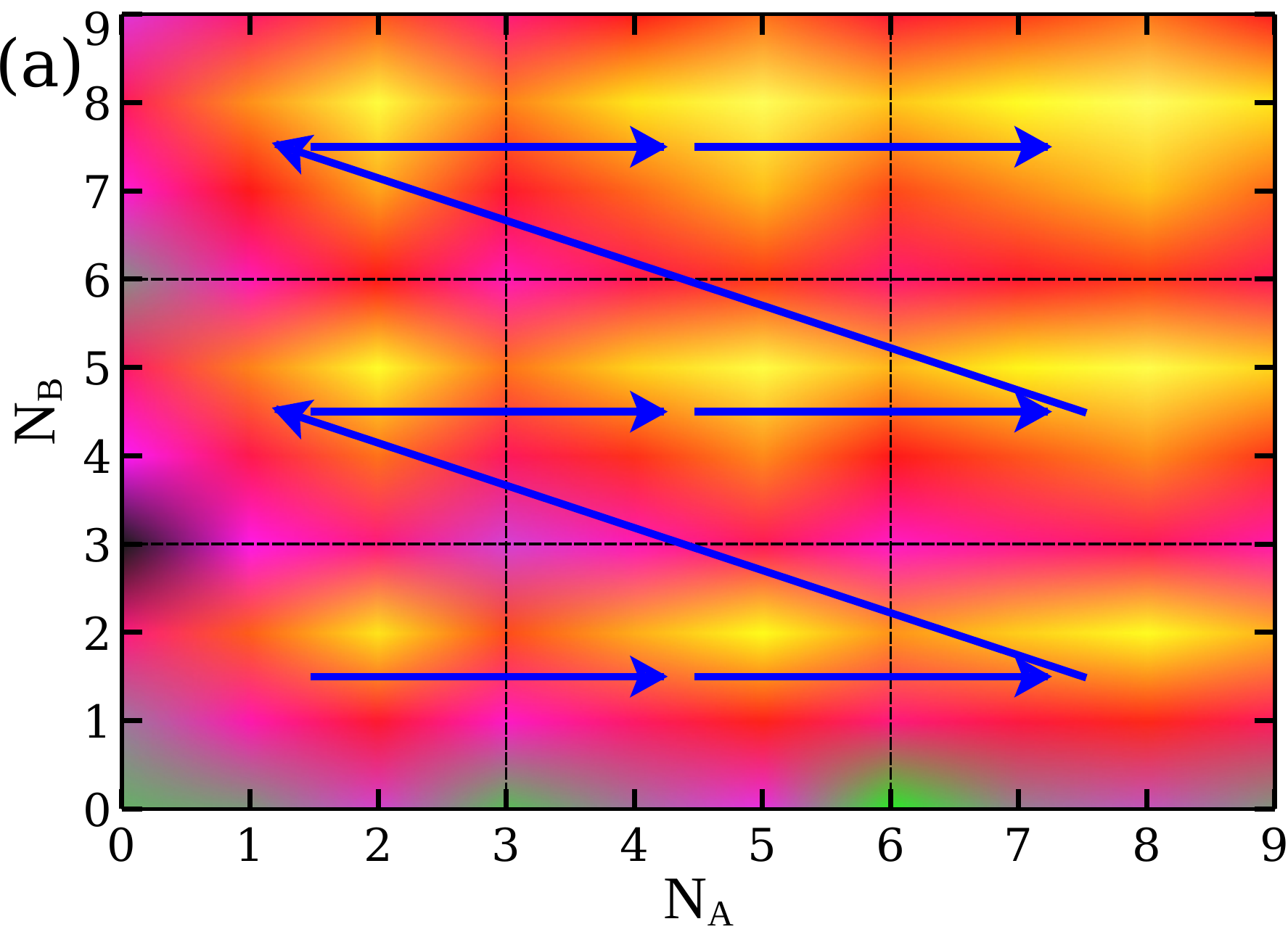}
\centering \includegraphics[width=1\linewidth]{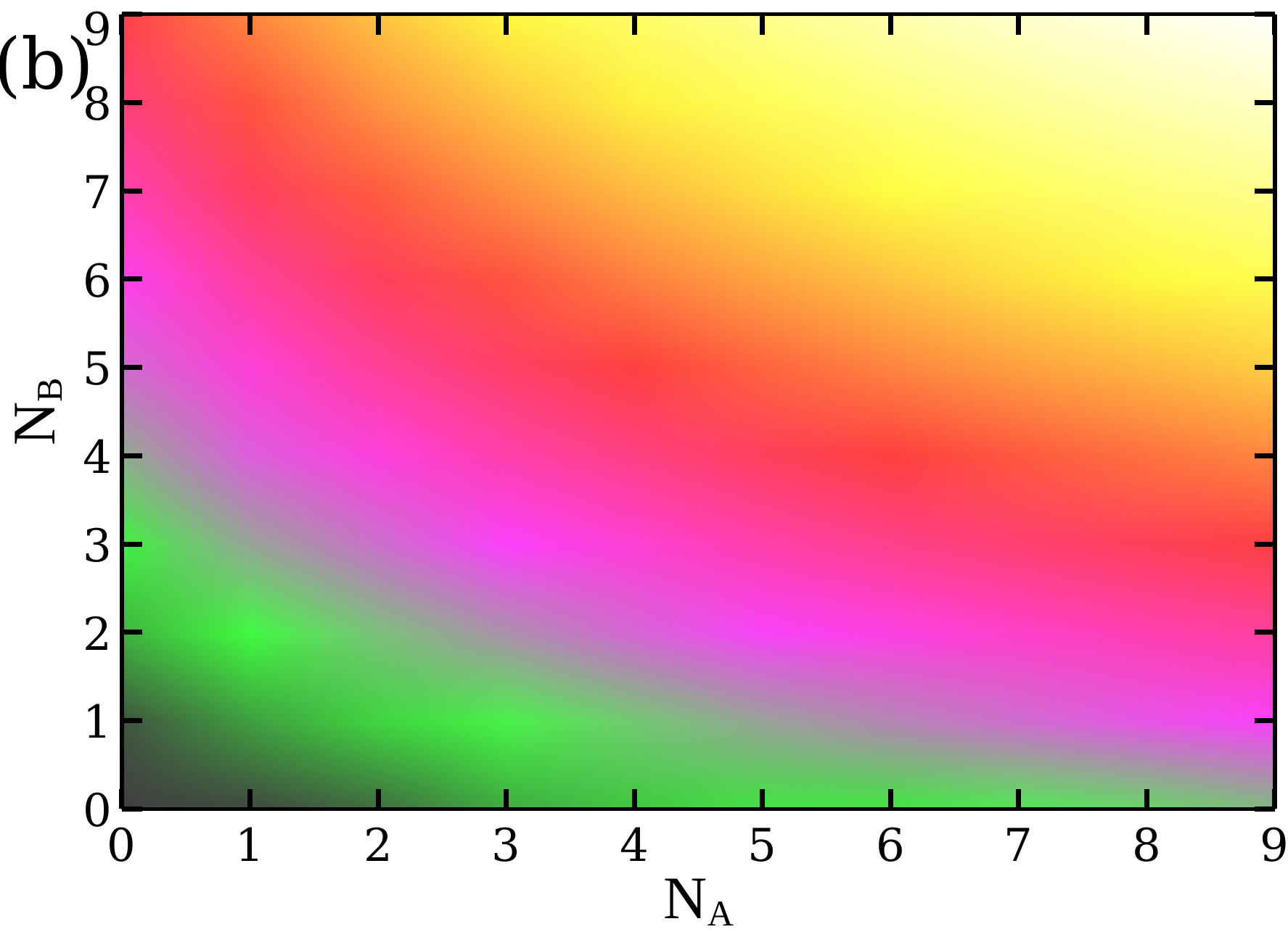}
\caption{Description of the scheme employed to reconstruct the complete density of states $p(N_A, N_B)$ for a binary mixture. (a) Raw output from the nine $3 \times 3$ windows with the lowest number of particles. The arrows show the order with which the windows are joined together. (b) The same data has been used to compute the final $p(N_A, N_B)$: there are no visible boundaries between the windows. The brighter the color, the higher the value of $p$. The curves have been smoothed out in order to increase readability.}
\label{fig:small_sus}
\end{figure}

There is no unique way of performing this operation, since the process of joining different windows can start from any window and follow any pattern. For the present study the scheme delivering the best results is the following. Since the system under study is a binary mixture, we compute the joint probability distribution $p(N_A, N_B)$, {\it i.e.,} we keep track of how many times the system has $N_A$ particles of type $A$ and $N_B$ particles of type $B$. 
We start with the $(0, 0)$ window and then begin to attach windows along the first species' direction, so that the second window is $(1, 0)$, the third is $(2, 0)$ and so on. Once the last window in the row has been joined, we start attaching another row of windows by joining the $(0, 1)$ window. This procedure is schematically shown in Figure~\ref{fig:small_sus}.

At low temperatures, {\it i.e.,} when the numerical noise increases, using Eq.~\eqref{eq:join} on the simulation data may result in histograms which cannot be reliably reweighted at all the required chemical potentials. This happens because the main contributions to the factors $b^w$ are the largest $p^w(N_A, N_B)$ values and, since reweighting to very different chemical potentials moves the signal to the less precisely-attached windows, the quality of the resulting histogram deteriorates. To overcome this difficulty we first reweight each $p^w$ and then join them together by means of Eq.~\eqref{eq:join}.

The simulation output $p(N_A, N_B)$, computed at fixed $T$, $\mu_A$ and $\mu_B$, can be evaluated at different chemical potentials $\mu_A'$, $\mu_B'$ by histogram reweighting, {\it i.e.,}

\begin{equation}
p'(N_A, N_B) \propto p(N_A, N_B) \mathrm{e}^{N_A(\beta \mu_A' - \beta \mu_A)} \mathrm{e}^{N_B(\beta \mu_B' - \beta \mu_B)},
\end{equation}

\noindent
where $\beta = 1/(kT)$, here $k$ is the Boltzmann constant. 
Therefore, $p(N_A, N_B)$ encodes all the information on the system in the whole investigated $N_A$, $N_B$ plane at fixed $T$. 

In order to obtain information on the phase behaviour of the mixture we employ the following criterion: we reweight $p(N_A, N_B)$ at a certain $\mu_A$ and then we tune $\mu_B$ until $p(N_A, N_B)$ is double-peaked, with the area below the two peaks being equal. If no such $\mu_B$ value can be found, then we are out of the coexisting region for the chosen $(T, \mu_A)$ values. If the equal area condition is fulfilled, the total free-energy density profile can be split up as a sum of two contributions $p_1(N_A, N_B)$ and $p_2(N_A, N_B)$, one for each phase. In the systems studied here, this is done by making a cut in $N_A$, $N_B$ plane at fixed $N_A=N_A^m$, where $N_A^m$ is the position of the fitted minimum of the $p(N_A) = \sum_{N_B=1}^{N_B^{\mathrm max}} p(N_A, N_B)$ curve. The number of particles of species $i$ in the phase $j$, $\langle N_i^j \rangle$ is then computed by taking an average over the appropriate particle number distribution, {\it i.e.,}

\begin{equation}
\langle N_i^j \rangle = \frac{\sum_{n_A=0}^{N_A^{\mathrm{max}}} \sum_{n_B=0}^{N_B^{\mathrm{max}}} n_i p^j(N_A, N_B)}{\sum_{n_A=0}^{N_A^{\mathrm{max}}} \sum_{n_B=0}^{N_B^{\mathrm{max}}} p^j(N_A, N_B)}.
\end{equation}

The quantities can, in turn, be used to compute compositions $x^{(j)}=\langle N_B^{(j)}\rangle/(\langle N_A^{(j)}+N_B^{(j)}\rangle)$ and densities $\rho^{(j)} = (\langle N_A^{(j)}+N_B^{(j)}\rangle) / V$.

The pressure $P$ can be computed by considering that $p(0, 0) = \mathrm{e}^{-\beta PV}$ is the grand-canonical partition function, and hence~\cite{shen_gc_pressure}

\begin{equation}
\label{eq:P}
P = -\frac{kT}{V} \log(p(0,0)).
\end{equation}
\noindent

Similarly to what we do for pure systems, we compute the (pseudo-)critical points of mixtures by comparing the $p(M$), obtained by projecting the $p(N_A, N_B)$, to the Ising order-parameter distribution of the right dimensionality~\cite{ising_2D,ising_3d}, with the only difference being the choice of the order parameter~\cite{virnau_alkanes_mixtures}, defined as

\begin{equation}
\label{eq:M_mixtures}
\Delta M \propto \rho_A + c \rho_B,
\end{equation}
\noindent
where $c$ is a fitting parameter which depends on temperature. Note that, unlike to what we do in pure systems, we do not store any information on the energy of the system, and hence we do not perform any temperature reweighting.

In the rest of the article, the superscript $^{2D}$ ($^{3D}$) is used to refer to quantities associated to the $2D$ ($3D$) model.

The simulation box sizes $L^{2D} = 16.9$ and $L^{3D} = 10$ are kept fixed throughout this work. We do not perform any finite-size scaling study and hence we compute only pseudo-critical parameters. For the sake of brevity, in the following we use \textit{critical} instead of \textit{pseudo-critical} when referring to these quantities.

\subsection{Theory}
We also investigate the patchy colloidal mixture theoretically by means of Wertheim's first order perturbation theory. A detailed description of the original theory can be found in Refs.~\cite{wertheim1,wertheim2,wertheim3,wertheim4}. Here we briefly quote the results and set the notation for Wertheim's theory extended to binary mixtures~\cite{mixtures_lisbona_1,macdowell_wertheim,virnau_alkanes_mixtures}. The Helmholtz free energy per particle of the mixture is:
\begin{eqnarray}
f_{H}=F_H/N=f_{ref}+f_b,
\end{eqnarray}
where $N=N_A+N_B$ is the total number of particles, $f_{ref}$ is the free energy per particle of the reference fluid of hard spheres (HSs) in 3D or hard disks (HDs) in 2D, and $f_b$ is the bonding free energy per particle. As usual we write $f_{ref}$ as the sum of ideal-gas and excess terms: $f_{ref}=f_{id}+f_{ex}$. The ideal-gas free energy is given (exactly) by
\begin{equation}
\beta f_{id}(\eta,x^{(i)})=\ln\eta-1+\sum_{i=A,B}x^{(i)}\ln(x^{(i)} {\cal V}_i),
\end{equation}
where ${\cal V}_i$ is the (irrelevant) thermal volume, $x^{(i)}=N_i/N$ is the molar fraction of species $i=\{A,B\}$, and $\eta=\eta_A+\eta_B$ is the total packing fraction ($\eta=v_s\rho$, with $\rho$ the total number density and $v_s=\pi/6\sigma^3$ the volume of a HS in 3D or $v_s=\pi/4\sigma^2$ the area of a HD in 2D). The excess part accounts for the excluded volume interactions between the monomers. Both species have the same size and hence we can approximate the excess part by the well-know Carnahan-Starling equation of state for hard spheres in the 3D mixture:
\begin{equation}
\beta f_{ex}(\eta)=\frac{4\eta-3\eta^2}{(1-\eta)^2}\;\text{ (3D) },
\end{equation}
and use the Henderson \cite{doi:10.1080/00268977500102511} equation of state for hard disks in the 2D mixtures:
\begin{equation}
\beta f_{ex}(\eta)=-\frac78\ln(1-\eta)+\frac98\frac\eta{(1-\eta)}\;\text{ (2D) }.
\end{equation}
The bonding free energy is approximated by Wertheim's thermodynamic first-order perturbation theory
\begin{equation}
\beta f_b=\langle M \rangle\left(\ln X_u-\frac{X_u}2+\frac12\right),
\end{equation}
where $X_u$ is the probability that one site is {\it not} bonded and
\begin{equation}
\langle M \rangle=x^{(A)}M^{(A)}+x^{(B)}M^{(B)},
\end{equation}
is the average number of patches per particle in the mixture ($M^{(A)}$ and $M^{(B)}$ are the number of patches of species $A$ and $B$ respectively). The probability of finding an unbonded patch is related to the total density, molar fractions and absolute temperature through the law of mass action  
\begin{equation}
X_u = 1 - \eta X_u^2 \Delta_u \langle M \rangle
\end{equation}
The bond between two patches is characterized by $\Delta_u$. Using the Kern-Frenkel potential, we find:
\begin{equation}
\Delta_{u}=\frac1{v_s}\int_{v_{b}}g({\bf r})\left[\exp(\beta\epsilon)-1\right]d{\bf r},\label{deltau}
\end{equation}
where $g({\bf r})$ is the radial distribution function of the reference fluid of HS or HD, and the integral is calculated over the volume (area) of a bond, $v_b$. If $v_b$ is small enough, we can approximate the radial distribution function by its contact value, $g_c(\eta)$. Under this assumption the Eq. \eqref{deltau} simplifies to
\begin{equation}
\Delta_{u}=\frac{v_b}{v_s}g_c(\eta)\left[\exp(\beta\epsilon)-1\right].
\end{equation}
The contact value of the radial distribution function is
\begin{equation}
g_c(\eta)=\frac{1-\eta/2}{(1-\eta^3)}\;\text{ (3D)}
\end{equation}
for hard spheres and
\begin{equation}
g_c(\eta)=\frac1{1-\eta}+\frac9{16}\frac\eta{(1-\eta)^2}\;\text{ (2D)}
\end{equation}
for hard disks. The bonding volume (area) is related to the depth and range of the potential. For the three dimensional mixtures we find
\begin{equation}
v_b=\pi(1-\cos\theta_{max})^2((\sigma+\delta)^3-\sigma)/3=0.00269\sigma^3\;\text{ (3D)},
\end{equation} 
and for the two dimensional case
\begin{equation}
v_b=\frac{\theta_{max}^2}{\pi}\left(\left[\sigma+\delta\right]^2-\sigma^2\right)=0.00418\sigma^2\;\text{ (2D)}.
\end{equation}
Finally, we obtain the equilibrium properties of the mixture by minimising (at a fixed pressure, composition and temperature) the Gibbs free energy per particle $g_{G}(x,\rho,P,T) = P/\rho + f_H$ with respect to the total density. We use a standard Newton-Raphson method to minimise $g_G$. Coexisting points are located by a standard common-tangent construction on $g_{G}(x)$ at constant temperature and pressure. Critical points are computed by determining those states which satisfy the spinodal condition, $f_{vv}f_{xx}-(f_{xv})^2=0$. In addition, stability requires the vanishing of the third-order derivative in the direction of largest growth:
\begin{equation}
f_{vvv}-3f_{xxv}\left(\frac{f_{xv}}{f_{vv}}\right)+3f_{xvv}\left(\frac{f_{xv}}{f_{vv}}\right)^2-f_{vvv}\left(\frac{f_{xv}}{f_{vv}}\right)^3=0,
\end{equation}
where subscripts denote partial derivatives, {\it i.e.}, $f_{xv}$ is the second partial derivative of $f_H$ with respect to the reduced volume per particle $v\equiv 1/\eta$ and the composition $x$ (molar fraction of bi-functional particles) at constant temperature.

\section{Results}

\subsection{Critical points}

In both investigated binary mixtures, particles of species $B$ are bi-functional and therefore do not exhibit any gas-liquid phase separation~\cite{bian,BianchiJCP07}. On the other hand, particles of species $A$ have a higher valence and exhibit a regular gas-liquid phase separation at low densities and temperatures~\cite{bian,FoffiKern}. 

\begin{figure}
\centering \includegraphics[width=1\linewidth]{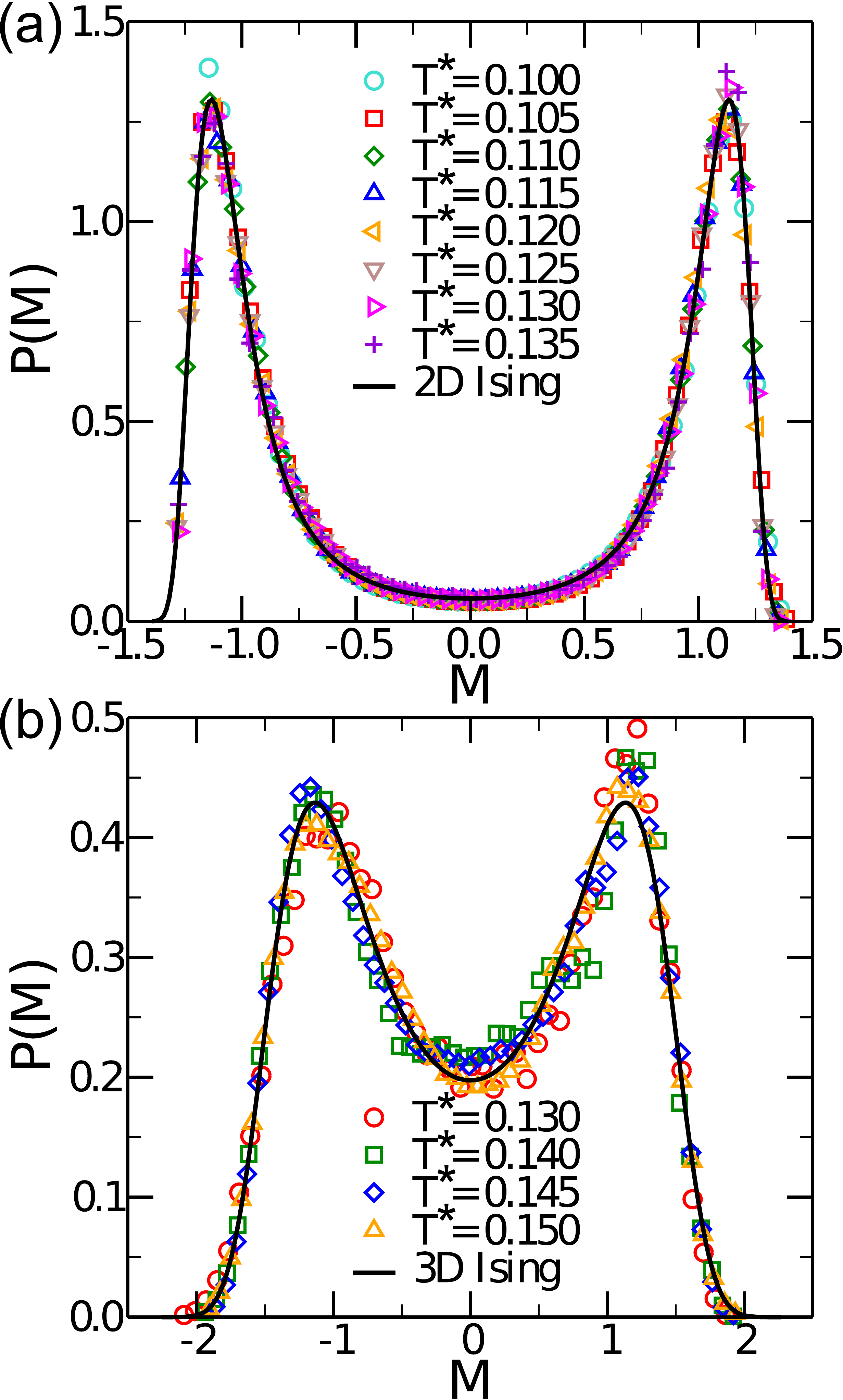}
\caption{Best fits to the Ising order parameter distribution for  mixtures of two- and three-functional patchy particles in 2D (a) and two- and tetra-functional patchy particles in 3D (b). Open symbols: Monte Carlo simulation at different reduced temperatures $T^*=kT/\epsilon$. Solid black line: Ising order parameter distribution.}
\label{fig:fits}
\end{figure}

We start by computing the pseudo-critical parameters of the pure systems, {\it i.e.,} of the tetra-functional model in 3D and of the three-functional model in 2D. We then move down in temperature and compute the phase boundaries on the whole $N_A$, $N_B$ plane at fixed temperature.
Figure~\ref{fig:fits} shows the best fits to the 2D and 3D Ising order parameter distribution for all the binary systems studied in this work. Data are less scattered in the 2D case but, despite the noise, the resulting fits are reliable at all temperatures. 

\begin{figure}
\centering \includegraphics[width=1\linewidth]{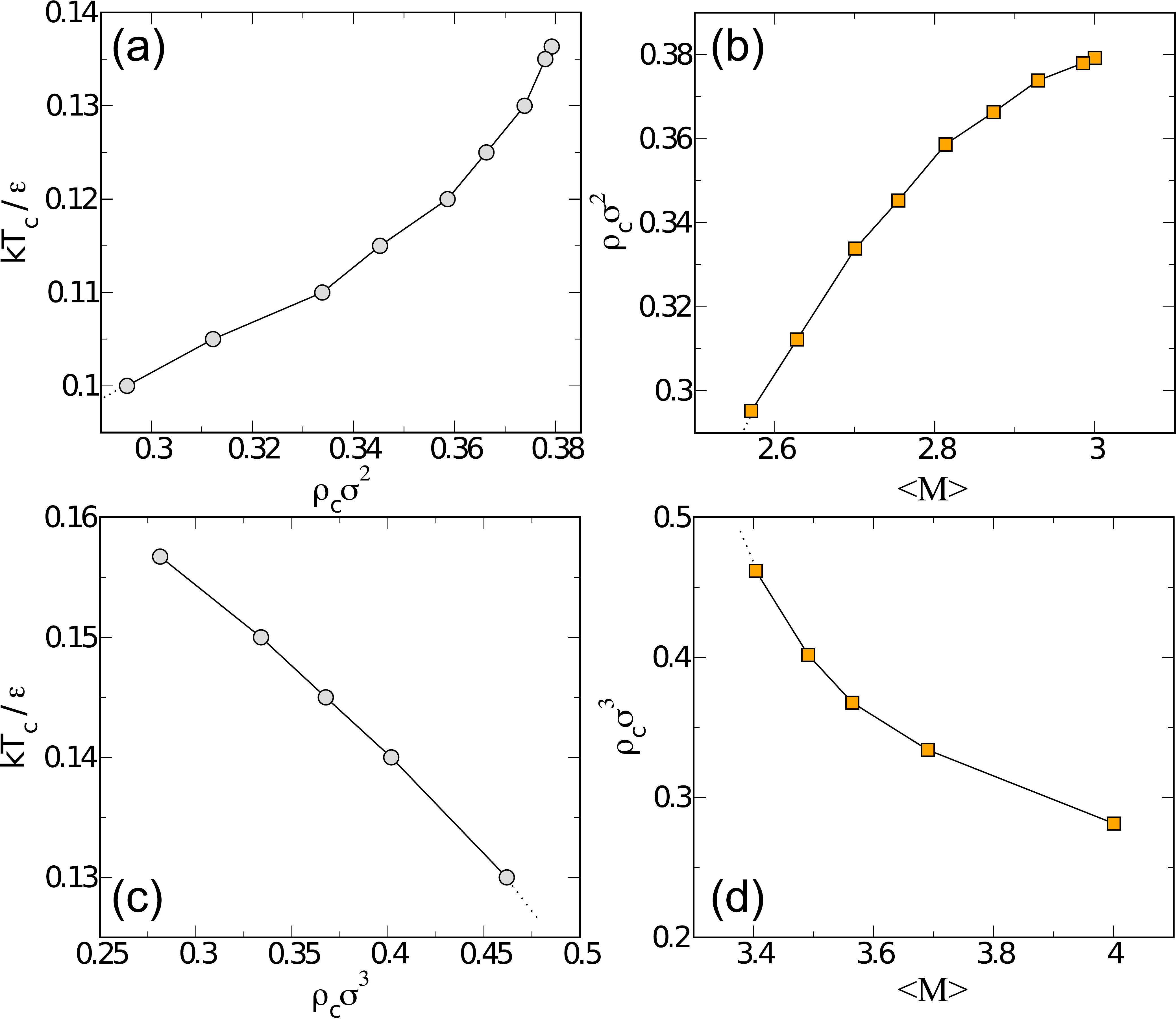}
\caption{Critical parameters for (a)-(b) a mixture of hard disks with two and three patches and for (c)-(d) a mixture of hard spheres with two and four patches, according to Monte Carlo simulation. Panels (a) and (c): reduced critical temperature as a function of the critical density. Panels (b) and (d): critical density as a function of the average number of patches per particle at the critical point.}
\label{fig:rhoc_Tc}
\end{figure}

From the $p(N_A, N_B)$ reweighted at criticality we can extract all the critical parameters of the mixtures, such as critical composition $x_c$, critical density $\rho_c$, critical pressure $P_c$ and average valence $\langle M \rangle$. Figure~\ref{fig:rhoc_Tc} shows the critical parameters of all the computed critical points according to Monte Carlo simulation. In a mixture of two- and three-functional patchy particles in 2D, see panel (a), the critical density goes down as the critical temperature decreases. Moreover, $\rho_c$ is a monotonically increasing function of $\langle M \rangle$, as shown in panel (b). By contrast, the mixtures of two- and tetra-functional patchy particles in 3D display a $\rho_c$ drop as the average valence is increased, see panel (d), as predicted by Wertheim's theory~\cite{mixtures_lisbona_1}. This growth is reflected in the behaviour of $\rho_c$ versus $T_c$, panel (c), which is monotonically decreasing in the investigated range of temperatures. In both systems the dependence on $T_c$ of the critical pressure $P_c$ (not shown) follows that of the critical density in this range of temperatures: $P_c$ increases with $T_c$ in 2D and decreases in 3D.

\begin{figure}
\centering \includegraphics[width=1.\linewidth]{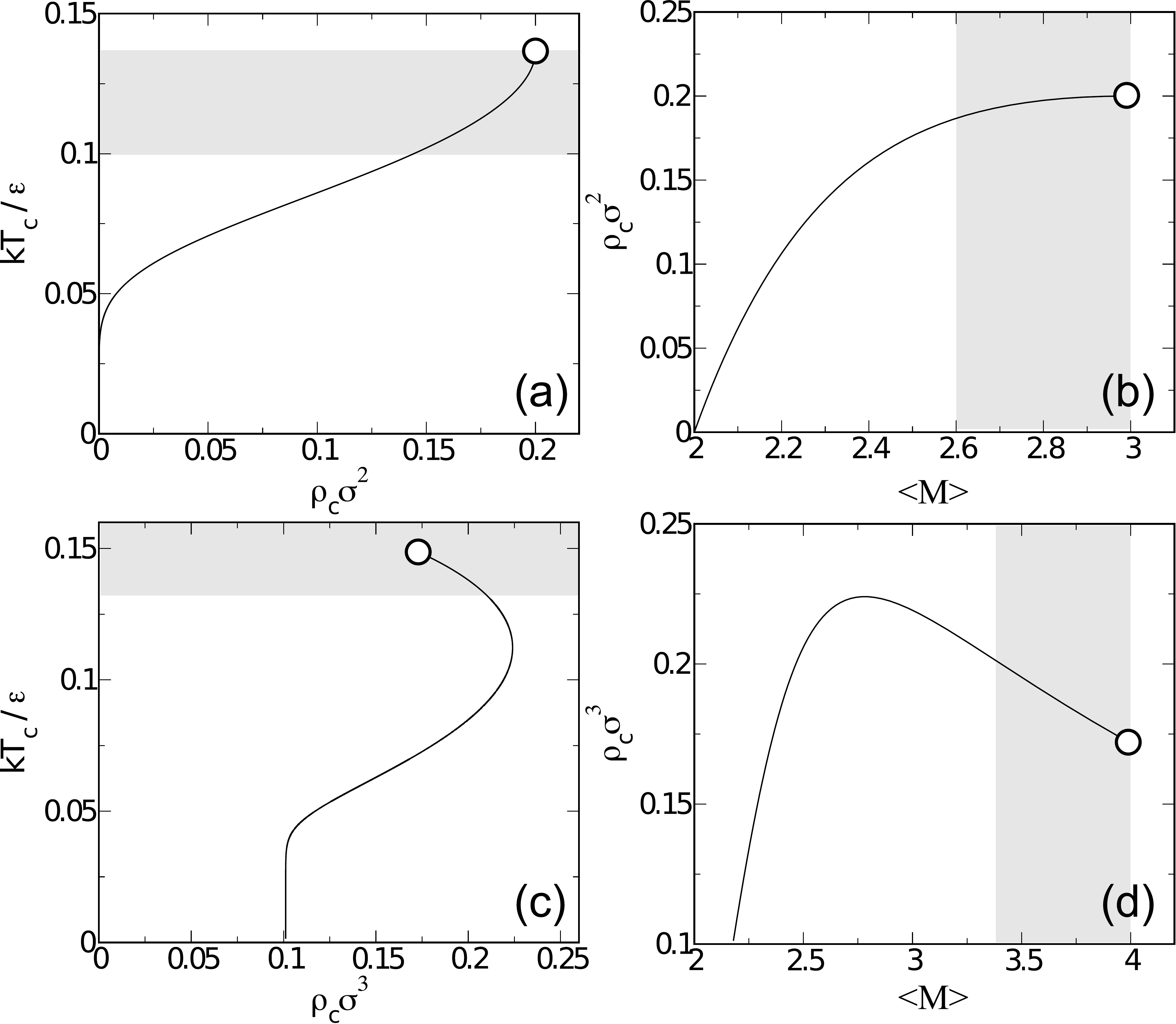}
\caption{Critical parameters for (a)-(b) a mixture of hard disks with two and three patches and (c)-(d) for a mixture of hard spheres with two and four patches, according to Wertheim's theory. Panels (a) and (c): reduced critical temperature as a function of the critical density. Panels (b) and (d): critical density as a function of the average number of patches per particle at the critical point. The empty circles indicate the position of the critical point for a pure fluid of three- (tetra-) functional particles in 2D (3D). The region explored via MC simulations is shaded in grey.} 
\label{fig:rhoc_Tc_teo}
\end{figure}

The results from Wertheim's theory are depicted in Figure~\ref{fig:rhoc_Tc_teo}. The theory predicts the same behaviour as Monte Carlo simulation for all the critical parameters and both types of mixtures considered here. The agreement is quantitative only for the critical temperature and the average valence at the critical point. Wertheim's first order perturbation theory does not include the formation of closed loops of patchy particles. As a result, the critical density is underestimated. This discrepancy may also arise from the different nature of the critical phenomenon, which is mean-field in Wertheim's theory and Ising in simulations.

In 2D (bi- and three-functional particles) and in the limit of zero temperature (note that using Wertheim's theory we can investigate the whole range of temperatures) the critical density vanishes, panel (a), and the valence at the critical point tends asymptotically to two, panel (b). Mixtures of bi- and tetra-functional particles in 3D behave differently, see panels (c) and (d). The theory predicts a reentrant behaviour for the critical density occurring for $kT/\epsilon\lesssim 0.12$, {\it i.e.,} for temperatures which are lower than those studied here by Monte Carlo simulations. At very low temperatures the critical density tends asymptotically to a value different from zero, and the average valence at the critical point is always higher than two. The critical pressure (not shown) also shows a non-monotonic behaviour upon lowering $T$. It first increases near the critical temperature of the pure tetra-functional fluid, and then decreases. For mixtures of bi- and five-functional particles (not shown) in 2D and in 3D the critical density increases monotonically as $T$ is lowered~\cite{mixtures_lisbona_1}.

\begin{figure}
\centering \includegraphics[width=1\linewidth]{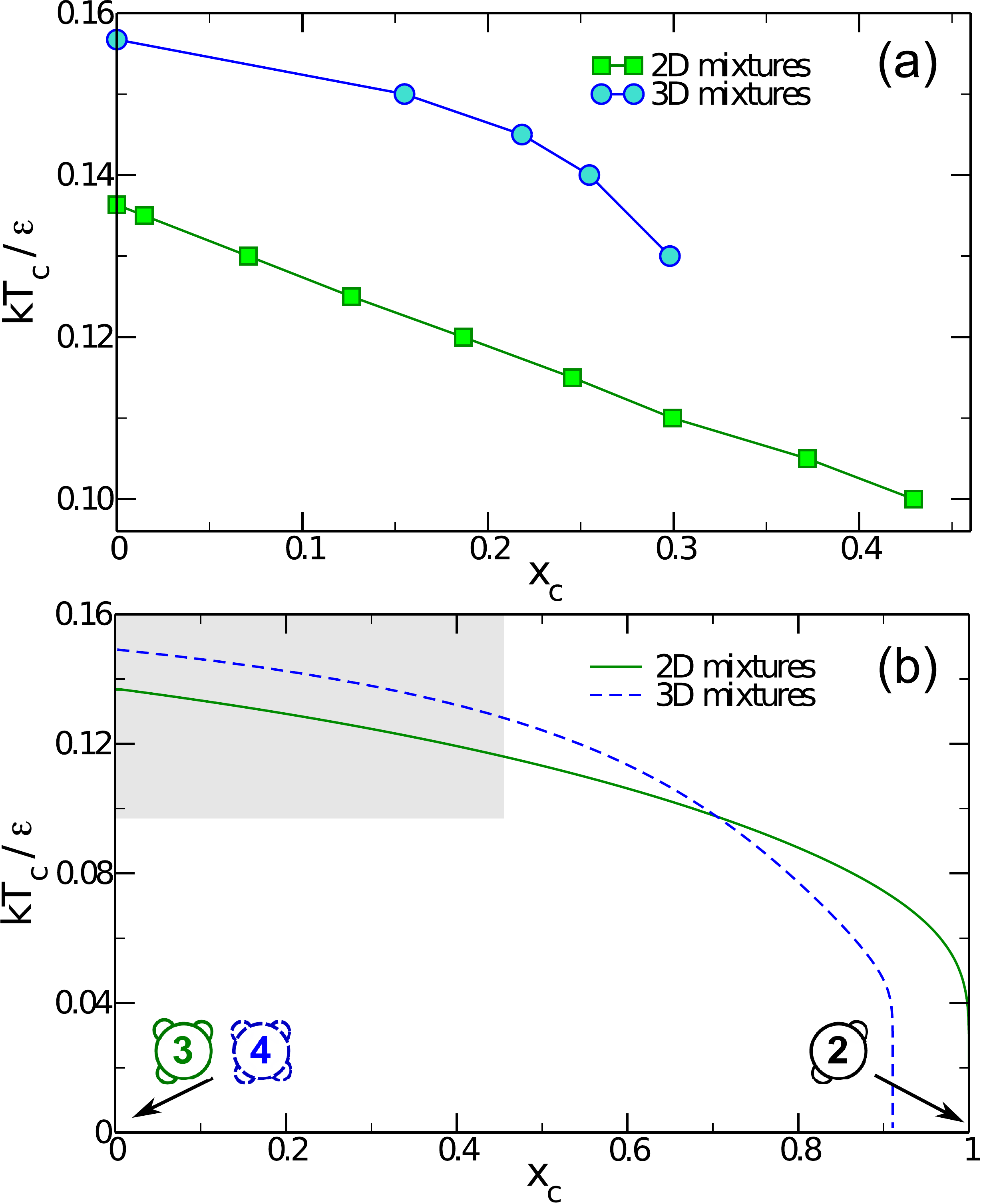}
\caption{Reduced critical temperature, $kT_c/\epsilon$, as a function of composition (i.e. fraction of bi-functional particles) at the critical point, $x_c$,  for mixtures of bi- and three-functional patchy particles in 2D and bi- and tetra-functional patchy particles in 3D. (a) Results from Monte Carlo simulation. (b) Results according to Wertheim's theory. The region investigated with MC simulation is shaded in grey.}
\label{fig:xc_Tc}
\end{figure}

To conclude this section on critical points, we compute the critical composition $x_c$, {\it i.e.,} the fraction of bi-functional particles (species $B$) at criticality, as a function of $T_c$. This quantity, shown in Figure~\ref{fig:xc_Tc}, is monotonically decreasing for both types of mixtures, as predicted by the theory. Therefore,  as the temperature is lowered, the composition of the fluid at criticality tends to favour more and more the bi-functional particles. At very low temperatures $x_c$ tends asymptotically to one in mixtures of bi- and three-functional particles (2D), and to $\approx0.9$ in mixtures of two- and tetra-functional particles (3D).

\subsection{Coexistence region}
\label{subsec:coexistence}

\begin{figure}
\centering \includegraphics[width=1\linewidth]{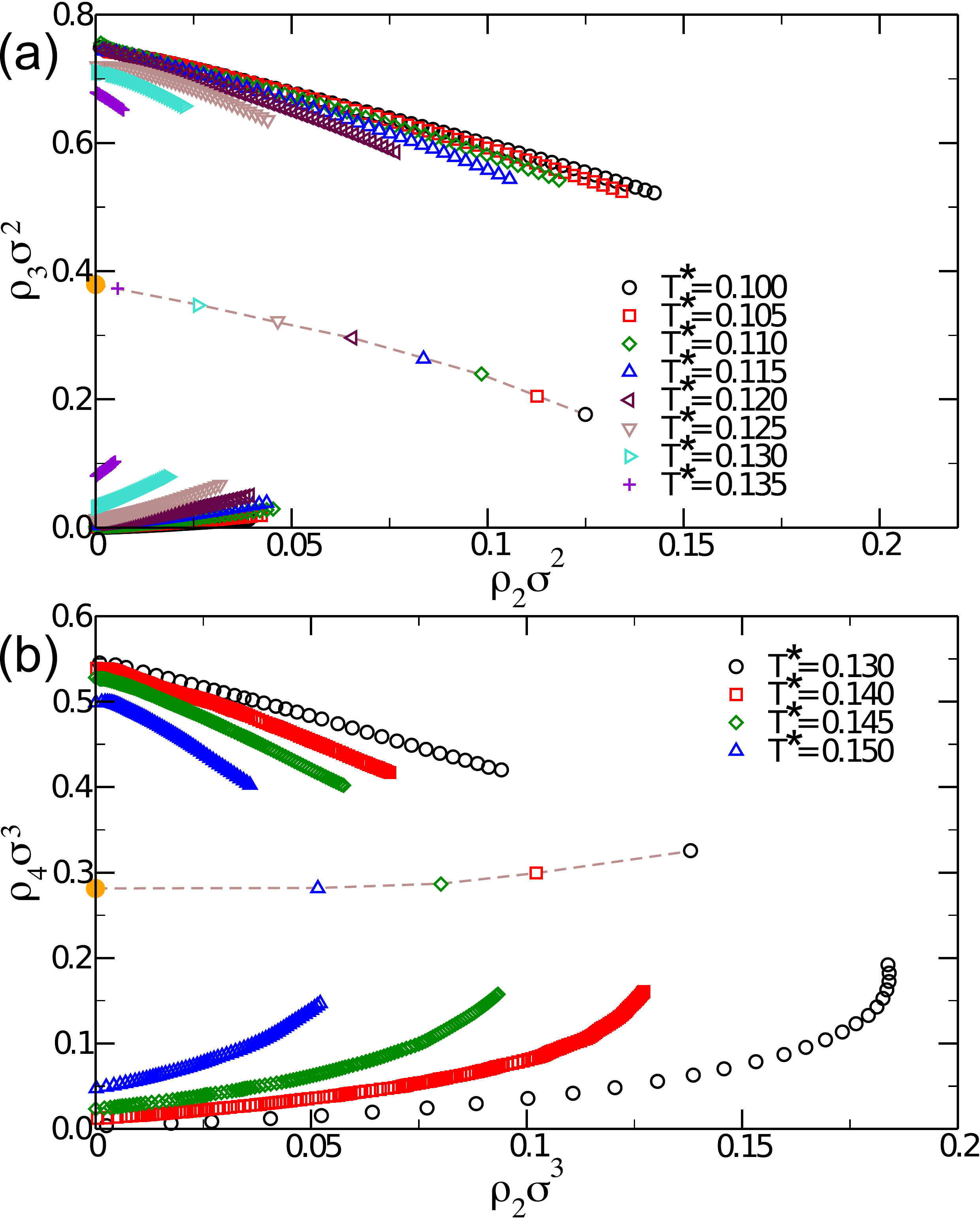}
\caption{Projection of the phase diagram of the (a) 2D binary mixture in the $\rho_2$, $\rho_3$ plane, and (b) 3D binary mixture in the $\rho_2$, $\rho_4$ plane according to MC simulation. Symbols: densities of the coexisting phases for different reduced temperatures $T^*=kT/\epsilon$. The symbols connected by the dashed line mark the positions of critical points. The orange filled circle marks the position of the critical point associated to the pure system of three (four) patches in 2D (3D), occurring at $T_c^{(3)}=0.136$ ($T_c^{(4)}=0.157$). Note that tie lines connecting two coexisting points (not shown) are not vertical lines in this plane.}
\label{fig:3D_tot_densities}
\end{figure}

We proceed to analyse the phase boundaries. For the mixtures considered here, upon lowering the temperature the instability region is first encountered at the critical temperature of the pure system, namely $T_c^{(A)}$. If we project the three-dimensional phase diagram onto the $\rho_A$, $\rho_B$ plane, then the phase-coexisting region is a point lying on the $\rho_B=0$ axis at $T=T_c^{(A)}$. If $T$ decreases, the instability region expands towards larger values of $\rho_B$, in line with the results for the critical parameters reported in the previous section. This $T$-dependence is shown in Figure~\ref{fig:3D_tot_densities}, which displays the phase boundaries and the computed critical points projected onto the $\rho_A$, $\rho_B$ plane for both types of mixtures. The temperature dependence in the explored $T$ range is qualitatively different in the two cases. 

The overall density of the low-density phase in the two dimensional $2-3$ mixture, see panel (a), does not change much upon changing $T$. The high-density phase branch, on the other hand, displays a rather strong $T$-dependence: as $T$ is lowered, it extends to larger values of $\rho_2$ and smaller values of $\rho_3$. Therefore, at all the investigated temperatures the phase transition has the character of an ordinary gas-liquid phase separation, with a rather broad separation in densities between the two phases.

Three dimensional $2-4$ mixtures, however, behave in a different manner, see panel (b). Even though the effect of $T$ on the liquid branch is similar to that observed in the 2D system, the phase originating from the gas-phase of the pure system does not remain confined at low densities but it moves at larger $\rho_2$ and $\rho_4$. Indeed, at the lowest temperature, the difference in overall density between the two coexisting phases is remarkably smaller than in the 2D case. In this case the phase transition has a stronger demixing component which tends to segregate the bi-functional particles in the low-density phase. 
This may be ascribed to a change in the balance between the entropy of mixing and the entropy of bonding as the number of patches $n$ on particles of species $A$ changes. When two particles of species $A$ form a bond, the result is a two-particle cluster with $2(n-1)$ sites available for bonding. By contrast, a cluster of two particles of species $A$ and $B$ has $n$ available sites. Therefore, the difference between the number of available bonding sites on the two types of clusters is $n-2$ sites. That is, as $n$ increases, forming a cluster of dissimilar species is less favourable entropically than having a cluster made up of particles of the same species. The gain in the entropy of mixing, on the other hand, remains the same. As a result, the tendency for phase separation increases with $n$. Ref.~\cite{mixtures_lisbona_1} contains a more detailed analysis on this topic.

\begin{figure}
\centering \includegraphics[width=1\linewidth]{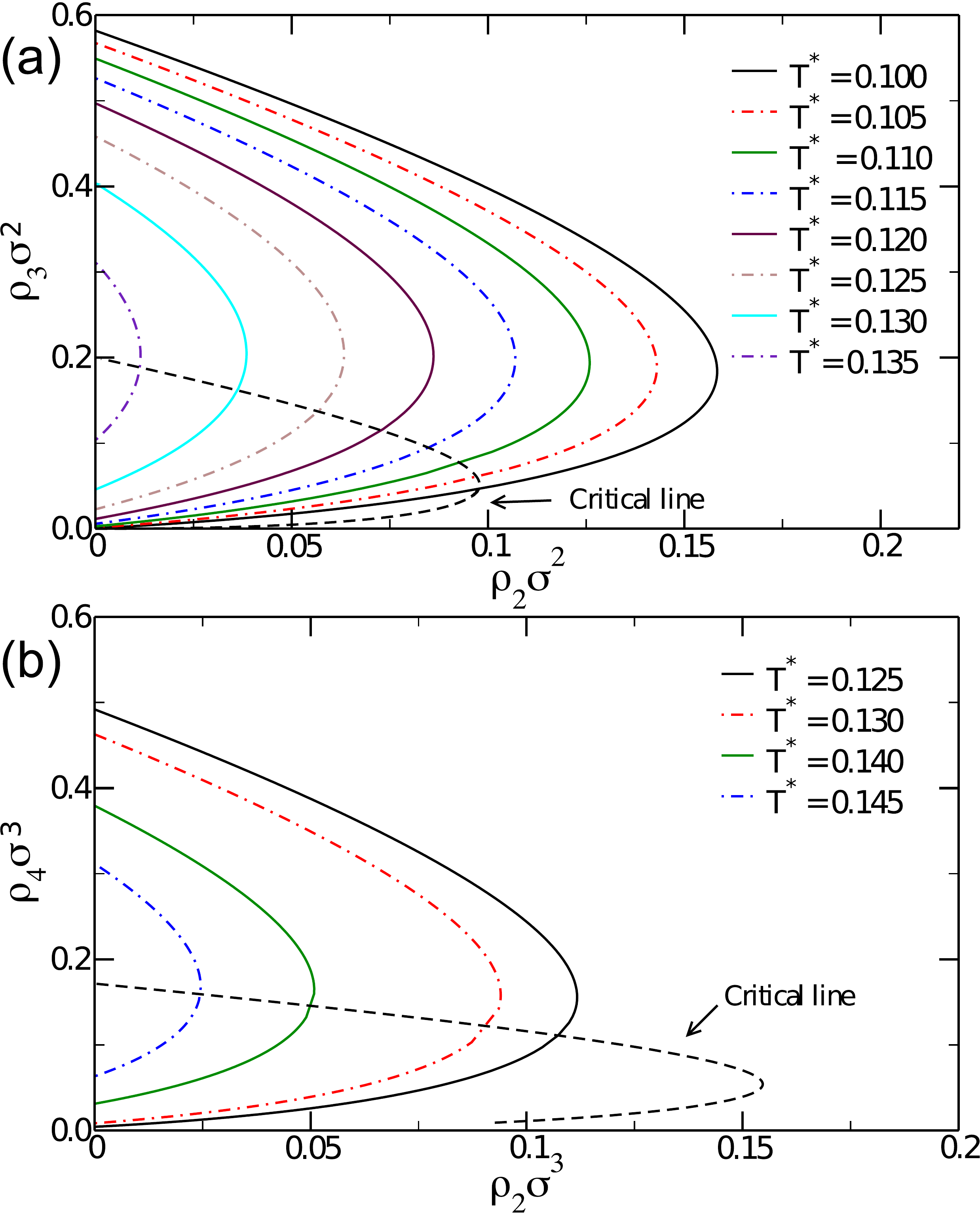}
\caption{Projection of the phase diagram of the (a) 2D binary mixture in the $\rho_2$, $\rho_3$ plane, and (b) 3D binary mixture in the $\rho_2$, $\rho_4$ plane (b) according to Wertheim's theory for different reduced temperatures $T^*=kT/\epsilon$. The black dashed line is the line of critical points of the mixture. Tie lines are not vertical lines in this representation.}
\label{fig:3D_tot_densities_teo}
\end{figure}

\begin{figure}
\centering \includegraphics[width=1\linewidth]{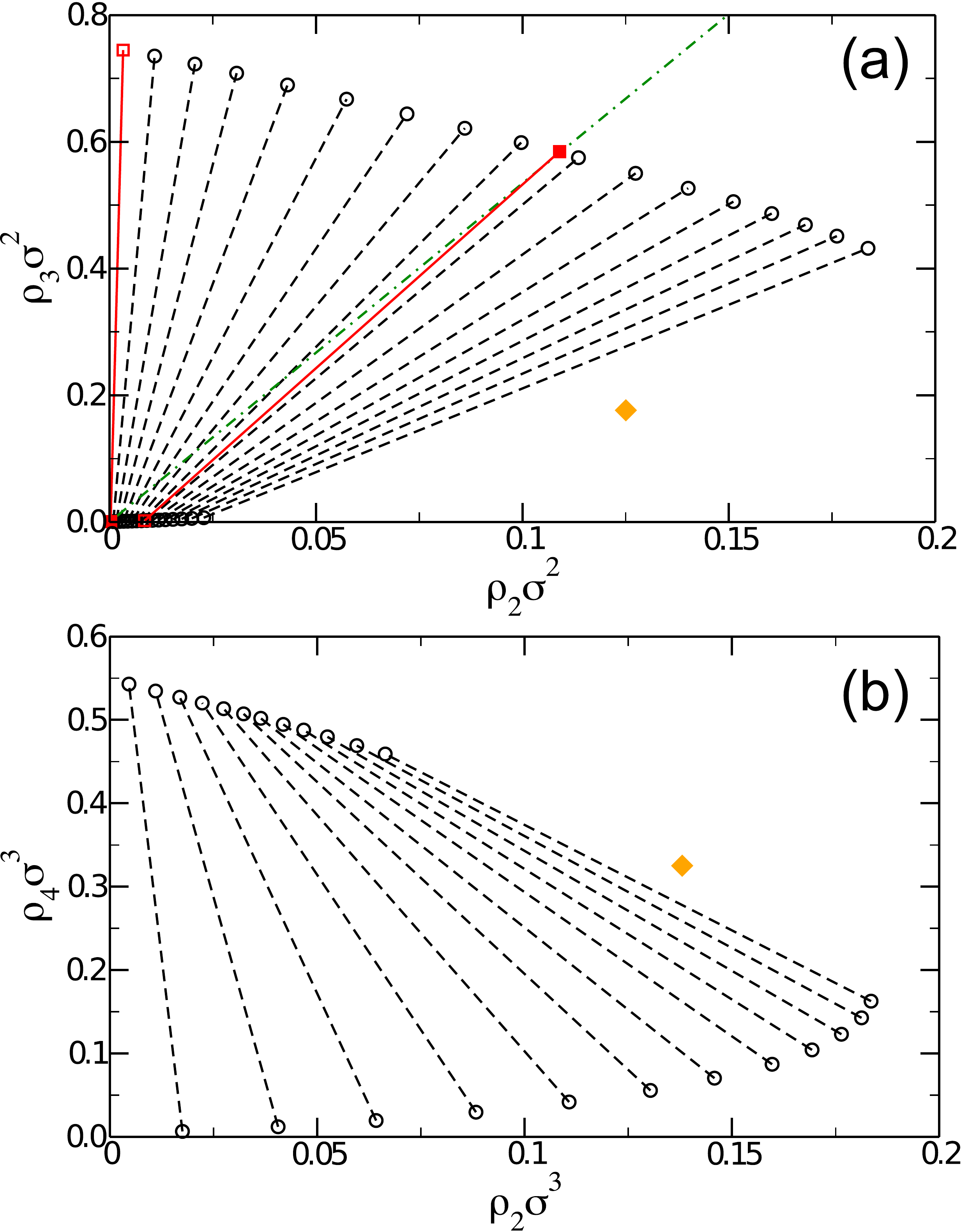}
\caption{Tie lines for the lowest-temperature system in (a) 2D and (b) 3D (dashed lines). Orange diamonds pinpoint critical points. The shadow-cloud construction is highlighted in panel(a). The dash-dotted line is the dilution line at fixed composition $x=0.181$. It crosses the coexistence region in two points which are part of the cloud curve (filled red squares). Their coexisting partners are part of the shadow cloud (open red squares).}
\label{fig:tie_lines_sim}
\end{figure}

\begin{figure}
\centering \includegraphics[width=1\linewidth]{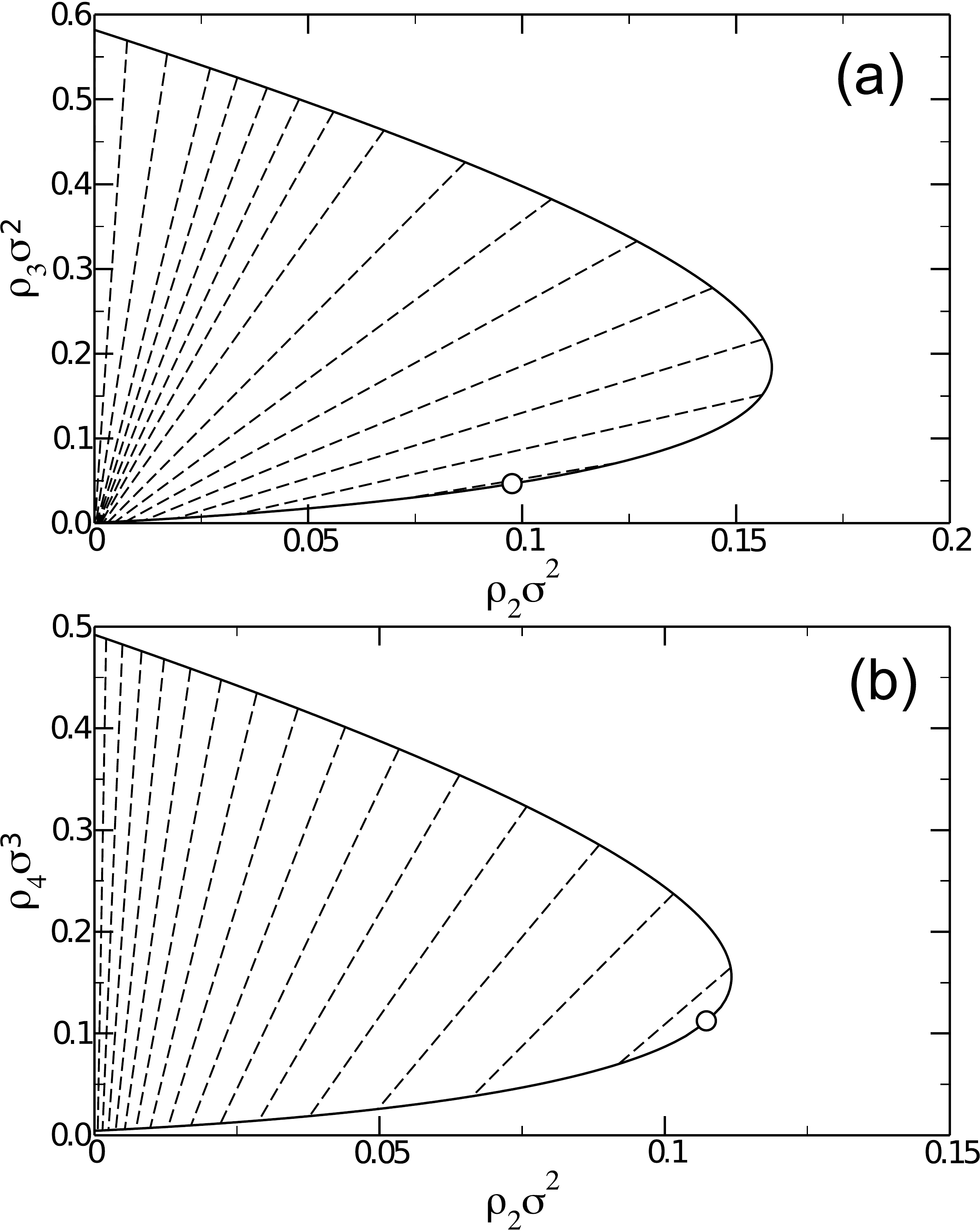}
\caption{Theoretical results for tie lines at the lowest-temperature system in (a) 2D and (b) 3D (dashed lines). Open circles mark the positions of critical points.}
\label{fig:tie_lines_teo}
\end{figure}

The same results according to Wertheim's theory are presented in Figure~\ref{fig:3D_tot_densities_teo}. As expected, the theory underestimates the coexisting densities both in 2D and 3D. As shown in the previous section, the 2D critical line tends asymptotically to $\rho_2=0$ and $\rho_3=0$, which makes it possible to find coexisting phases at arbitrarily low densities. By contrast, the 3D critical line ends at finite densities $\rho_2$ and $\rho_4$. 
In addition, in the 2D mixture the theory captures the overall behaviour of the system. In particular, the decrease of $\rho_2$ and $\rho_3$ at criticality, observed in simulations, is well reproduced. In the 3D case, on the other hand, there is a qualitative difference between numerical and theoretical results. As the system is cooled down, there is an increase of the critical $\rho_2$ and $\rho_4$ as computed in simulations, while the theory predicts a decrease of the critical $\rho_4$. This difference is associated to a different slope of tie lines, i.e. of the lines connecting the two coexisting phases. Tie lines for the lowest-$T$ systems are shown in Figure~\ref{fig:tie_lines_sim} (simulations data) and in Figure~\ref{fig:tie_lines_teo} (theoretical results).

If tie lines have positive slopes, the coexisting points follow the same trend as constant-composition lines which, in this representation, are straight lines with zero intercept. This means that high-density phases always have a higher density of both species than the coexisting low-density phases. In the systems studied here, this is true for the 2D numerical and theoretical results and for the theoretical 3D mixture. By contrast, the numerical 3D system exhibits tie lines with negative slope: the gas-like phases always have more bi-functional particles than their respective liquid-like phases. In this regard, the theory fails to capture the demixing nature of the phase separation occurring in the 3D system as observed in simulations. On passing, we note that Wertheim theory predicts tie lines with negative slopes in mixtures of particles with two and six patches (now shown).

\begin{figure}
\centering \includegraphics[width=1\linewidth]{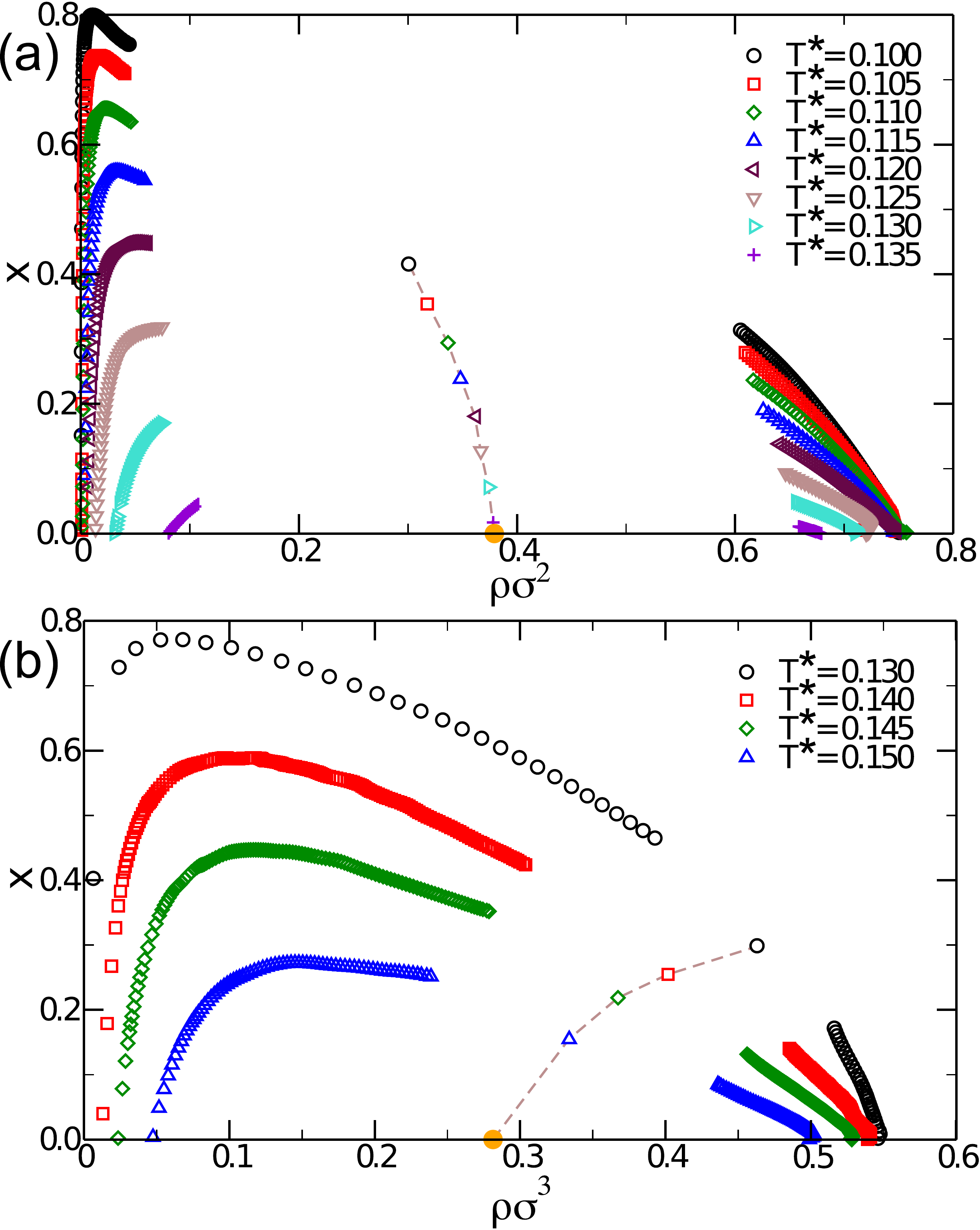}
\caption{Projection of the phase diagram of the (a) 2D binary mixture and the (b) 2D binary mixture in the $\rho$, $x$ plane according to MC simulation. Symbols: densities of the coexisting phases for different reduced temperatures $T^*=kT/\epsilon$. The symbols connected by the dashed line mark the positions of critical points. The orange point marks the position of the critical point of the associated pure system with three (four) patches in 2D (3D). Note that the tie lines are not vertical lines.}
\label{fig:3D_tot_x_rho}
\end{figure}

\begin{figure}
\centering \includegraphics[width=1\linewidth]{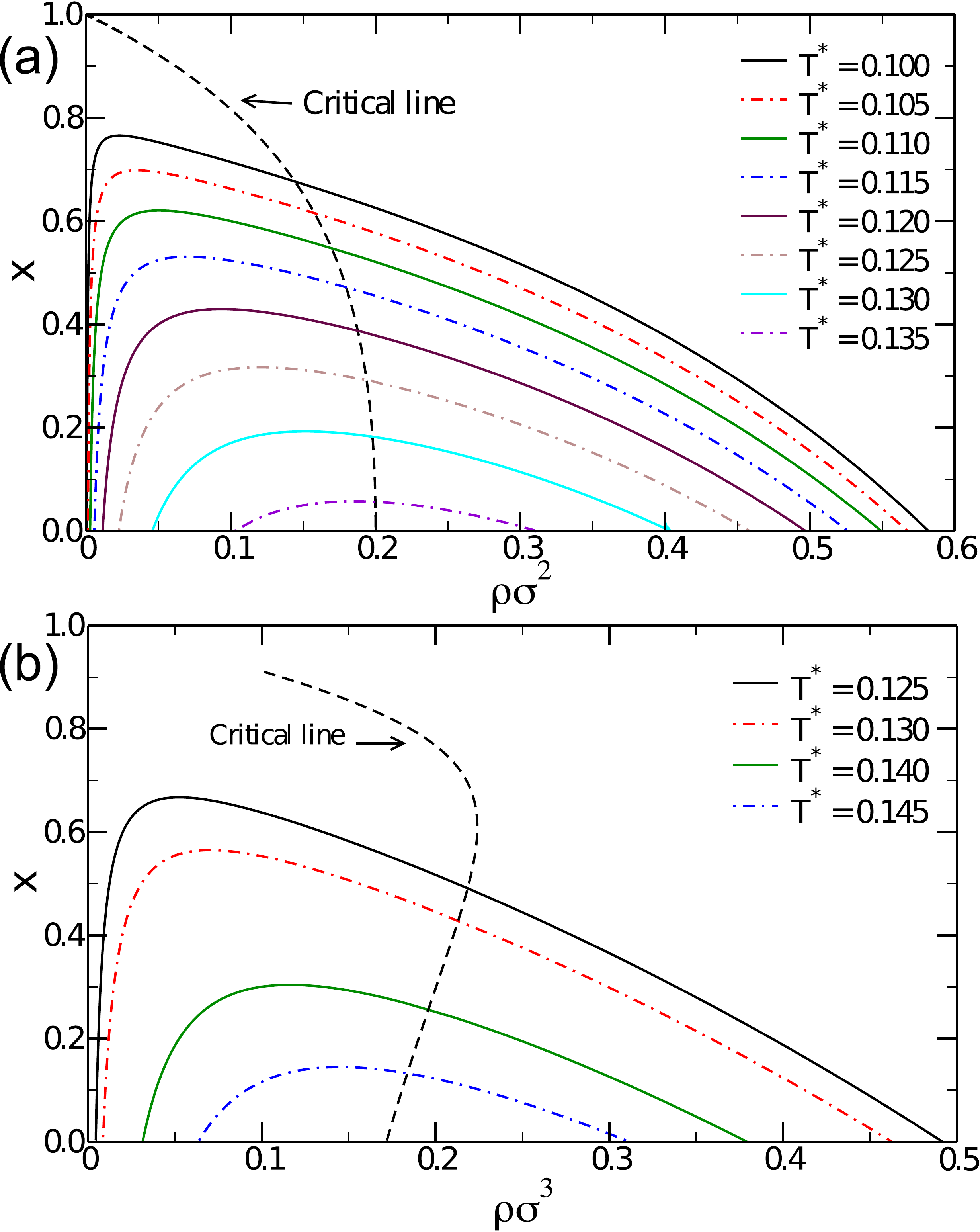}
\caption{Projection of the phase diagram of the (a) 3D binary mixture and the (b) 2D binary mixture in the $\rho$, $x$ plane for different reduced temperatures $T^*=kT/\epsilon$ according to Wertheim's theory. The black-dashed line is the line of critical points.}
\label{fig:3D_tot_x_rho_teo}
\end{figure}

A different representation can be constructed by projecting the three dimensional phase diagram onto the $\rho$, $x$ plane, {\it i.e.,} total density against composition (molar fraction of bi-functional particles). This is shown in Figure~\ref{fig:3D_tot_x_rho} (Monte Carlo simulation) and in Figure~\ref{fig:3D_tot_x_rho_teo} (Wertheim's theory). This representation makes it clear that the increase of the critical density in the 3D systems is due to the contribution of the low-density phase, which moves to larger densities and compositions as the temperature is lowered in the vicinity of the critical temperature of the pure four-patches fluid. In the 2D system, on the other hand, the density range of the low-density phase does not change strongly with $T$, while the composition increases steadily. This results in the shift of the critical density to smaller values. 

This particular projection of the phase diagram hides some of the differences we previously noted between theory and simulations. Indeed, tie lines in this representation always have negative slope.

\subsection{Cloud and shadow curves}

A common bi-dimensional representation of the phase diagram of binary mixtures is done by making a cut at a fixed composition in the three-dimensional $\rho_A, \rho_B, T$ phase diagram. The intersection between the cutting plane and a fixed temperature plane is usually called dilution line~\cite{pol_mixtures}, because by following it the density of the system can be varied without changing the overall composition. The points at which the dilution line intersect the phase boundary are the so-called cloud points. Each of these points coexists with a infinitesimal amount of the other phase which has, in general, a different composition. These points are called shadow points. An exemplification of this procedure is given in Fig.~\ref{fig:tie_lines_sim}(a). The shadow and cloud curves are then constructed by plotting the sets of cloud and shadow points on the $T$, $\rho$ plane. Note that this cloud-shadow construction is a projection and therefore, while the cloud points have the same composition, the composition of the shadow points varies, {\it i.e.,} the tie lines are out-of-plane. The spinodal lies inside the cloud curve and the two lines touch at the critical point, which is always located at the intersection of the shadow and cloud curves~\cite{buzzacchi_pre,sollich_poly_review}.

\begin{figure}
\centering \includegraphics[width=1\linewidth]{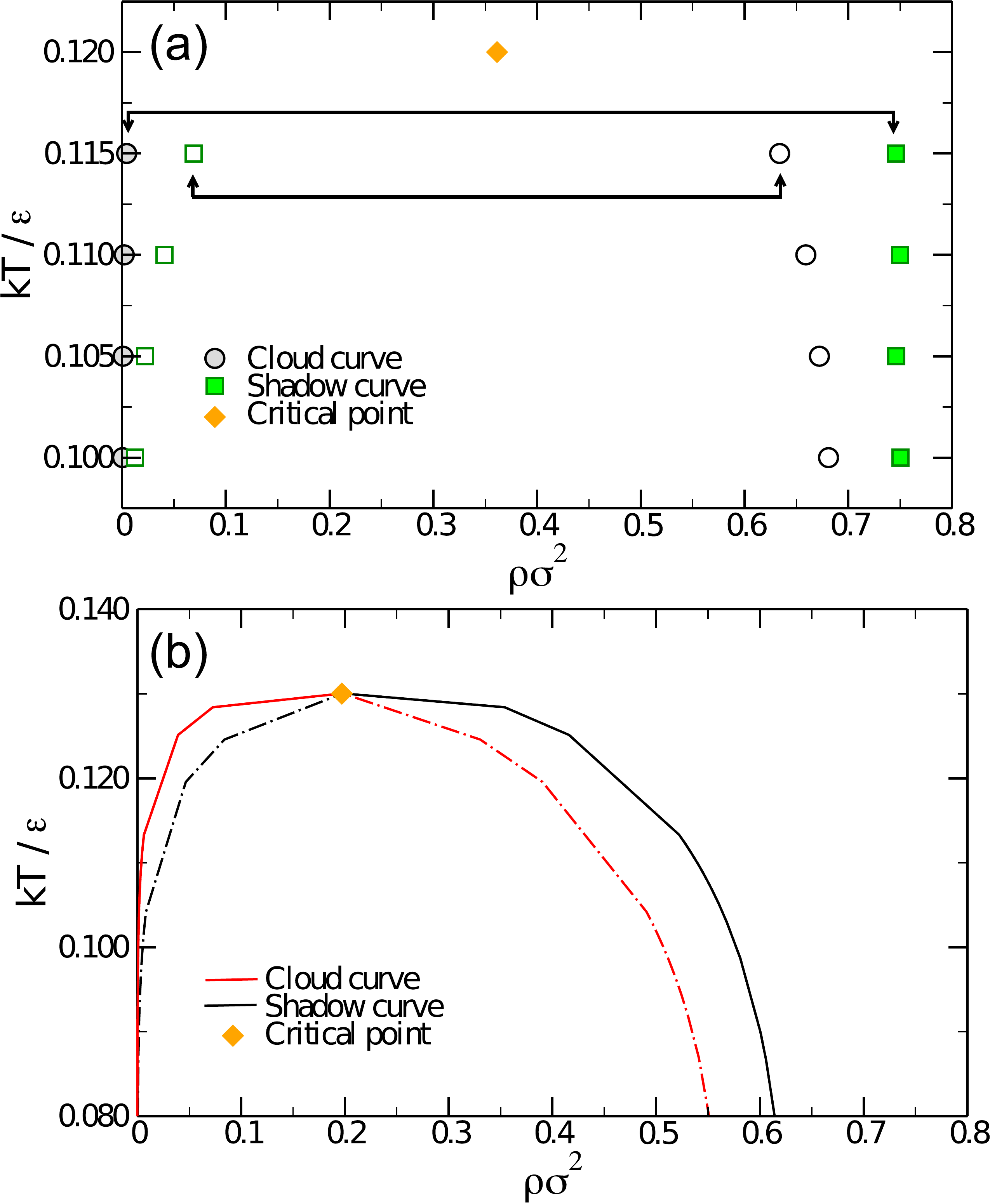}
\caption{(a) Cloud (black circles) and shadow (green squares) curves for the 2D binary system at fixed $x = 0.181$. Each point on the gas branch of the cloud curve (filled circles) is in coexistence with the point on the liquid branch of the shadow curve at the same temperature (filled squares), and vice versa (empty symbols). Lines with arrows point out the connection between cloud and shadow points. (b) Theoretical results for the same system at the same value of $x$. The orange diamonds signal the position of the critical point of the mixture having the same composition.}
\label{fig:2D_cloud}
\end{figure}

Figure~\ref{fig:2D_cloud} shows the cloud-shadow construction for the 2D system at a fixed composition $x = 0.181$. The orange diamond marks the position of the critical point at $kT/\epsilon=0.12$, which happens to have the same composition. In line with the results of Section~\ref{subsec:coexistence}, the curves resemble the usual bell-shaped gas-liquid coexisting regions observed in pure systems~\cite{hansen_verlet,flavio,lungo}. Both cloud and shadow curves have a low-density (high-density) phase which is in coexistence with a high-density (low-density) phase. In addition, all the coexisting branches have a monotonic $T$-dependence. Indeed, both the low-density cloud branch and the high-density shadow branch have nearly $T$-independent densities, whereas the densities of the low-density shadow branch and of the high-density cloud branch are monotonically decreasing and increasing, respectively, as in gas-liquid phase transitions of pure, simple fluids. All the cloud-shadow constructions computed for the 2D mixtures share the same qualitative features with the one shown here, regardless of $x$. Theoretical results, shown in Fig.~\ref{fig:2D_cloud}(b), are qualitatively in line with simulation data.

\begin{figure}
\centering \includegraphics[width=1\linewidth]{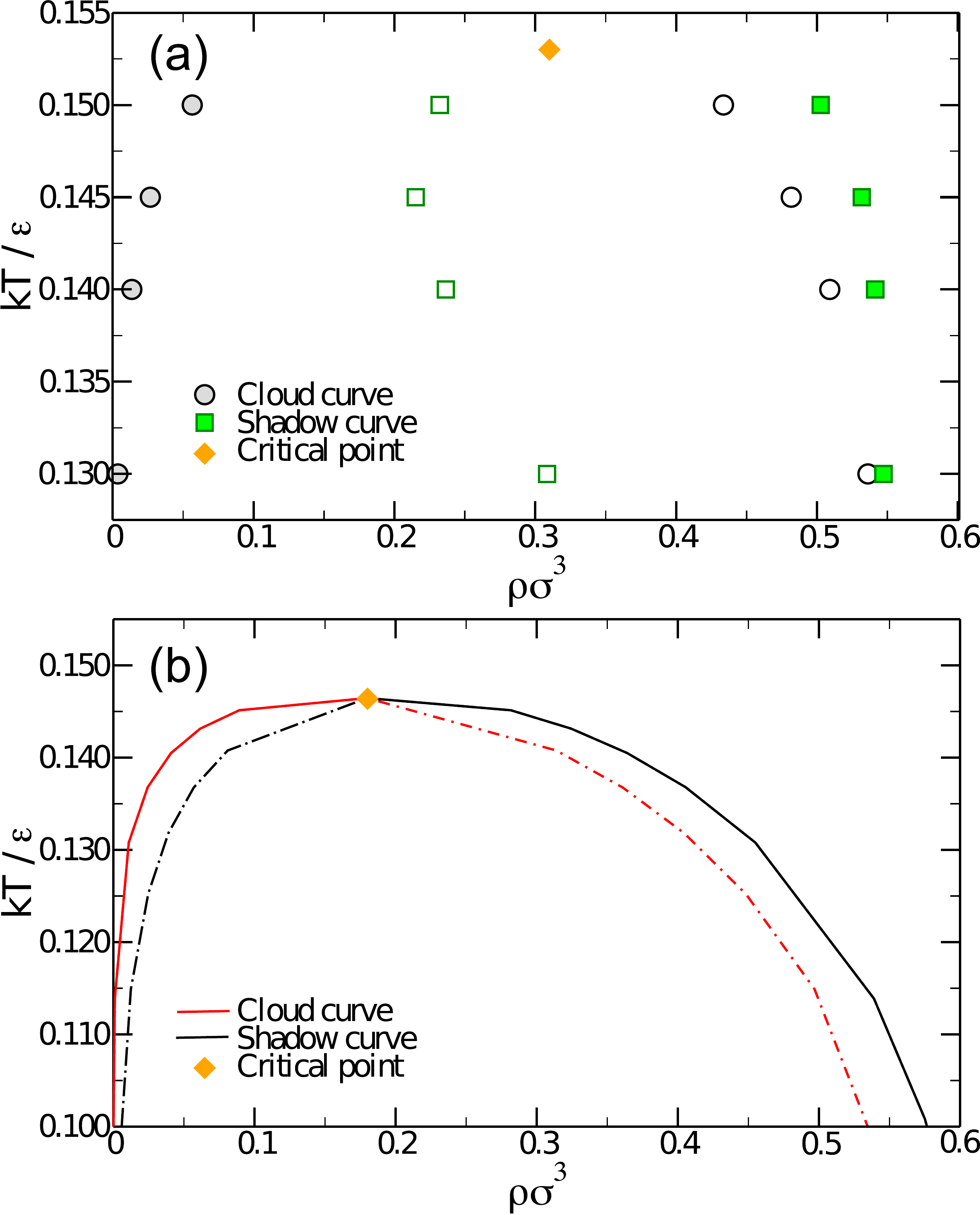}
\caption{(a) Cloud (black circles) and shadow (green squares) curves for the 3D binary system at fixed $x = 0.09$. The symbol filling has the same meaning as in Fig.~\ref{fig:2D_cloud}. The orange diamond signals the position of the critical point of the mixture having the same composition, extrapolated from the location of the nearby computed critical points. (b) Theoretical results for the same system at the same value of $x$.}
\label{fig:3D_cloud}
\end{figure}

The $4-2$ mixture, as previously noted, behaves in a different way. Figure~\ref{fig:3D_cloud} shows the cloud and shadow curves for a mixture with a fixed composition $x = 0.09$. The whole cloud curve has a monotonic $T$-dependence and its behaviour is qualitatively similar to the 2D case. The shadow curve, on the other hand, exhibits several remarkable features. First of all, the density of the low-density branch is rather high, being half-way between the two branches of the cloud curve and very close to the critical density. Moreover, its temperature dependence is non-monotonic, first decreasing and increasing upon lowering $T$. The high-density branch seems to be non-monotonic as well and, at a slightly lower temperature than the lowest investigated $T$, a crossing between the cloud and the shadow high-density branches is expected. We ascribe the lacking of this feature in the theoretical curves, shown in Fig.~\ref{fig:3D_cloud}(b), to the demixing character of the transition.

\section{Conclusions}

We have extended the successive umbrella sampling method~\cite{sus} to binary mixtures in oder to simulate the full bulk phase diagram. The method consists in dividing the simulation cell into a set of overlapping windows of variable size, which is set by the number of particles of each species allowed in a particular window. Each window is then sampled by means of grand canonical Monte Carlo simulations, counting the number of times that different microstates, characterized by the number of particles of each species, appears in each window. This gives the probability distribution of microstates, and thus the desired thermodynamic properties of the system. We have tested the validity of the new method by computing the phase behaviour of patchy colloidal mixtures. We have compared the results with the theoretical predictions of Wertheim's first order perturbation theory.

We have analysed two types of colloidal mixtures: bi- and three-functional patchy particles in two dimensions, and bi- and tetra-functional patchy particles in three dimensions. In the first case, bi- and three-functional patchy colloids in 2D, simulation and theory are in excellent agreement. The agreement is quantitative for those variables that do not involve the density, such as the temperature or composition. The theory underestimates the density (a well know problem of Wertheim's first order perturbation theory) as it neglects the formation of closed loops. As a consequence the agreement between the predicted densities and those found in the simulations is only qualitative. The phase behaviour of these mixtures is similar to that of bi- and three-functional particles in three dimensions~\cite{bian,mixtures_lisbona_1}: the critical density vanishes as the critical temperature approaches zero, suppressing condensation and yielding an increasingly large region of phase space where empty liquids are stable. Thus, dimensionality does not change the topology of the phase diagram of these mixtures. This was to be expected since the energy and entropy of bonding dominate the behaviour of such mixtures and this is determined by the functionality of the particles, rather than by the spatial dimension.

Three dimensional mixtures of bi- and tetra-functional colloids, however, exhibit a different behaviour. The results of Wertheim’s theory predict that as the critical temperature decreases the critical density first increases, then decreases and tends asymptotically to a value, which is different from zero. The simulation confirms the initial increase of the critical density but the non-monotonic behaviour could not be confirmed by simulations as it occurs (according to the theory) at very low temperatures, a region not accessible by the current simulation techniques. Despite the qualitative description of the mixture’s critical behaviour, the theory fails to describe adequately the shadow and cloud curves for these mixture. The origin of this discrepancy may be related to the formation of closed loops of particles, neglected by the theory, which increase in number as the functionality of the particles increases.

In summary we have developed and tested a new simulation scheme to investigate the phase diagram of binary mixtures of patchy colloidal particles. The new method can be applied to a large variety of problems, such as the surface and confinement properties of patchy colloidal mixtures, or the study of the bulk and percolation properties of the recently predicted bicontinuous gels or bigels~\cite{goyal_bicontinuous_gels,delasheras_bicontinuous_gels,varrato2012arrested}. In mixtures of patchy particles these bigels may be equilibrium structures, when they occur in the empty liquid regime \cite{delasheras_bicontinuous_gels} or dynamically arrested structures when they occur inside the liquid-vapour or the liquid-liquid binodals, as in ordinary binary mixtures \cite{varrato2012arrested}. A detailed investigation of the connectivity and other physical properties of these structures, in and out of equilibrium, is bound to reveal novel features with potential applications.  

\section*{Acknowledgements}

L.R. and F.S. acknowledge support from ERC-226207-PATCHYCOLLOIDS and ITN-234810-COMPLOIDS. J.M.T. and M.M.T.G. acknowledge financial support from the Portuguese Foundation for Science and Technology (FCT) under Contracts Nos. PEstOE/FIS/UI0618/2011 and PTDC/FIS/098254/2008. We thank Nigel Wilding for fruitful discussions and for having provided the 2D Ising $P(M)$.

\end{document}